# Knowledge synthesis from 100 million biomedical documents augments the deep expression profiling of coronavirus receptors

AJ Venkatakrishnan[1], Arjun Puranik[1], Akash Anand[2], David Zemmour[1], Xiang Yao[3], Xiaoying Wu[3], Ramakrishna Chilaka[2], Dariusz K. Murakowski[1], Kristopher Standish[3], Bharathwaj Raghunathan[4], Tyler Wagner[1], Enrique Garcia-Rivera[1], Hugo Solomon[1], Abhinav Garg[2], Rakesh Barve[2], Anuli Anyanwu-Ofili[3], Najat Khan[3], Venky Soundararajan[1]*

1. nference, inc., One Main Street, Suite 400, East Arcade, Cambridge, MA 02142, USA.
2. nference Labs, pvt. ltd., Wind Tunnel Road, Murugesh Palaya, Bengaluru 560017, India.
3. Janssen Research & Development LLC, USA.
4. nference, 111 Peter St, Toronto, ON M5V 2G9, Canada.

* Address correspondence to venky@nference.net

# Abstract

The COVID-19 pandemic demands assimilation of all available biomedical knowledge to decode its mechanisms of pathogenicity and transmission. Despite the recent renaissance in unsupervised neural networks for decoding unstructured natural languages, a platform for the real-time synthesis of the exponentially growing biomedical literature and its comprehensive triangulation with deep omic insights is not available. Here, we present the nferX platform for dynamic inference from over 45 quadrillion possible conceptual associations extracted from unstructured biomedical text, and their triangulation with Single Cell RNA-sequencing based insights from over 25 tissues. Using this platform, we identify intersections between the pathologic manifestations of COVID-19 and the comprehensive expression profile of the SARS-CoV-2 receptor ACE2. We find that tongue keratinocytes, airway club cells, and ciliated cells are likely underappreciated targets of SARS-CoV-2 infection, in addition to type II pneumocytes and olfactory epithelial cells. We further identify mature small intestinal enterocytes as a possible hotspot of COVID-19 fecal-oral transmission, where an intriguing maturation-correlated transcriptional signature is shared between ACE2 and the other coronavirus receptors DPP4 (MERS-CoV) and ANPEP (α-coronavirus). This study demonstrates how a holistic data science platform can leverage unprecedented quantities of structured and unstructured publicly available data to accelerate the generation of impactful biological insights and hypotheses.

The nferX Platform Single-cell resource - https://academia.nferx.com/

## Introduction

Since December 2019, the SARS-CoV-2 virus has been rapidly spreading across the globe. The associated disease (COVID-19) has been declared a pandemic by the World Health Organization (WHO), with over 350,000 confirmed cases and 15,000 deaths across nearly every country as of March 23, 2020[1]. The constellation of symptoms, ranging from acute respiratory distress syndrome (ARDS) to gastrointestinal issues, is similar to that observed in the 2002 Severe Acute Respiratory Syndrome (SARS) epidemic and the 2012 Middle East respiratory syndrome (MERS) outbreak. SARS, MERS, and COVID-19 are all caused by *Coronaviruses* (CoV), deriving their name from the crown-like spike proteins protruding from the viral capsid surface. Coronavirus infection is driven by the attachment of the viral spike protein to specific human cell-surface receptors: ACE2 for SARS-CoV-2 and SARS-CoV[2–4], DPP4 for MERS-CoV[5] and ANPEP for specific α-coronaviruses[6]. In addition to these receptors, the protease activity of TMPRSS2 has also been implicated in viral entry[7,8].

In a recent clinical study of COVID-19 patients from China, 48% of the 191 infected patients studied had comorbidities such as hypertension and diabetes[9]. Epidemiological and clinical investigations on COVID-19 patients are also suggesting fecal viral shedding and gastrointestinal infection[10–12]. In the case of the earlier SARS epidemic, multiple organ damage involving lung, kidney and heart was reported[13]. The mechanisms by which various comorbidities impact the clinical course of infections and the reasons for the observed multi-organ phenotypes are still not well understood. Thus, there is an urgent need to conduct a comprehensive pan-tissue profiling of ACE2, the putative human receptor for SARS-CoV-2.

A deep profiling of ACE2 expression in the human body demands a platform that synthesizes biomedical insights encompassing multiple scales, modalities, and pathologies described across the scientific literature and various omics siloes. With the exponential growth of scientific (e.g. PubMed, preprints, grants), translational (e.g. clinicaltrials.gov), and other (e.g. patents) biomedical knowledge bases, a fundamental requirement is to recognize nuanced scientific phraseology and measure the strength of association between all possible pairs of such phrases. Such a holistic map of associations will provide insights into the knowledge harbored in the world's biomedical literature.

While unsupervised machine learning has been advanced to study the semantic relationships between word embeddings[14,15] and applied to the material science corpus[16], this has not been scaled-up to extract the "global context" of conceptual associations from the entirety of publicly available unstructured biomedical text. Additionally, a principled way of accounting for the distances between phrases captured from the ever-growing scientific literature has not been comprehensively researched to quantify the strength of "local context" between pairs of biological concepts. Given the propensity for irreproducible or erroneous scientific research[17], which reflects as truths, semi-truths, and falsities in the literature, any local or global signals extracted from this unstructured knowledge need to be seamlessly triangulated with deep biological insights emergent from various omics data silos.

The nferX software is a cloud-based platform that enables users to dynamically query the universe of possible conceptual associations from over 100 million biomedical documents, including the COVID-19 Open Research Dataset recently announced by the White House[18] (**Figure 1**). An unsupervised neural network is used to recognize and preserve complex biomedical phraseology as 300 million searchable tokens, beyond the simpler words that have generally been explored using higher dimensional word embeddings previously[14]. Our *local context* score is derived from pointwise mutual information content between pairs of these tokens and can be retrieved dynamically for over 45 quadrillion possible associations. Our *global context score* is derived using word2vec[14], as the cosine distance between 180 million word vectors projected in a 300 dimensional space (**Figure 1A, Figure S1**).

In order to assess the veracity of these conceptual associations derived from biomedical literature, it is absolutely essential to enable triangulation with structured data sources including gene and protein expression datasets. To address this need and empower the scientific community, we built a Single Cell resource which harnesses these local and global score metrics to enable seamless integration of literature-derived associations with the analysis of transcriptomes from over 1 million individual cells from over 25 human and mouse tissues (**Figure 1B**). Here we use this first-in-class resource to conduct a comprehensive expression profiling of ACE2 across host tissues and cell types and discuss how the observed expression patterns correlate with the pathogenicity and viral transmission shaping the ongoing COVID-19 pandemic (**Figure 1C**).

## Results

### ACE2 has higher expression levels in multiple cell types of the Gastrointestinal (GI) tract compared to respiratory cells

To systematically profile the transcriptional expression of ACE2 across tissues and cell types, we triangulated single cell RNAseq-based measurements with literature-derived signals to automatically delineate novel, emerging, and known expression patterns (**Figure 2B; Table S1**). This approach immediately highlights renal proximal tubular cells and small intestinal enterocytes among the cell types that most robustly express ACE2 (detection in >40% of cells). These cell types are also moderately to strongly associated with ACE2 in the literature. The strong intestinal ACE2 expression is particularly interesting given the emerging clinical reports of fecal shedding and persistence post-recovery which may reflect a fecal-oral transmission pattern[10–12].

Conversely, pancreatic PP cells (gamma cells), pancreatic alpha cells, and keratinocytes show similarly robust ACE2 expression but have not been strongly associated with ACE2 in the literature. This combination suggests either a biological novelty or an experimental artifact. We note that the strong ACE2 expression in pancreatic cell types is derived from only one murine study (**Figure S2A**[19]), while ACE2 expression is not observed in gamma or alpha cells from scRNA-seq of human pancreatic islets (**Figure S2B**[20–22]). While we cannot determine the validity of either observation, this example demonstrates how knowledge synthesis can automatically surface discordant biological signals for further evaluation.

Surprisingly, cells from respiratory tissues were notably absent among the populations with highest ACE2 expression by scRNA-seq (**Figure 2B**). This observation was corroborated by complementary gene expression analysis of over 250,000 bulk RNA-seq samples from GTEx[23,24] and the Gene Expression Omnibus (GEO) along with protein expression analysis from healthy tissue proteomics and immunohistochemistry (IHC) datasets[25–27], where lung and other respiratory tissues consistently show lower ACE2 expression compared to the digestive tract and kidney (**Figure S3A-D**). However, the respiratory transmission of COVID-19 along with the disease symptomatology and well-documented viral shedding in respiratory secretions[28] strongly indicates that SARS-CoV-2 indeed infects and replicates within these tissues. This would suggest that ACE2 is likely expressed in the respiratory epithelium, and so we prioritized the respiratory and digestive tracts for further knowledge synthesis-augmented scRNA-seq analysis.

We also applied the *Single Cell resource* to analyze several other human and mouse tissues including heart, adipose, liver, pancreas, blood, spleen, bone marrow, thymus, testis, prostate, bladder, ovary, uterus, placenta, brain, and retina. An example use case describing the functionalities of the Single Cell app and a summary of ACE2 expression across these tissues are given in the **Supplemental Text** and **Figures S4-19**.

## Club cells, ciliated cells, and pneumocytes are likely targets of SARS-CoV-2 in respiratory tract

Next, we classified 105 respiratory cell populations from eight independent studies based on their expression of and literature-derived associations to ACE2 (**Figure 3A**). Consistent with the low levels of ACE2 in respiratory tissues by bulk RNA-seq, proteomics, and IHC (**Figure S3A-D**), we found that ACE2 expression is detected in fewer than 10% of all cell types recovered from these studies. However, as mentioned above, we believe that even low ACE2 expression levels in these respiratory cells may be sufficient for COVID-19 pathogenesis.

We found that club cells (formerly known as Clara cells), were consistently among the highest-expressing respiratory cell types (**Figures 3A-B**). Literature-derived local and global scores suggest that this ACE2-club cell connection is underappreciated, with a few documents discussing these concepts together ([nferx link](nferx link)). We also found that ACE2 is detected in type II pneumocytes in multiple studies, although the percentage of expressing cells ranges from only 0.5-7% (**Figures 3A-B**). This relatively low expression, which may be deemed inconsequential if viewed in isolation, is strongly supported by knowledge synthesis that highlights an existing association between ACE2 and type II pneumocytes (**Figures 3A-B**). Indeed, multiple studies have demonstrated ACE2 expression in these cells[29–35]. Further, ACE2 expression in bulk RNAseq of GTEx lung samples (n = 578) is strongly correlated to markers of type II pneumocytes, with all seven surfactant protein-encoding genes among the top 4% of transcriptional correlations to ACE2 (out of the ~19,000 genes expressed at > 1 TPM in GTEx lung samples; hypergeometric p-value = $1.1 \times 10^{-10}$) (**Figure S20**).

Our scRNAseq analysis also shows that ACE2 is expressed in small fractions of ciliated airway cells and epithelial cells of the nasal cavity (**Figures 3A-B**). While no staining is observed for ACE2 in nasopharynx samples from the Human Protein Atlas (HPA) IHC dataset (**Figure S21**), a previous IHC study did report the staining of ACE2 in nasal and oral mucosa and the nasopharynx[33]. This expression is consistent with the high SARS-CoV-2 viral loads detected in nasal swab samples[28]. Intriguingly, mild degeneration of olfactory epithelium was observed in an immunosuppressed animal model infected with SARS-CoV[36]. These observations are correlated with emerging reports of anosmia/hyposmia (loss of smell) in otherwise asymptomatic COVID-19 patients from South Korea and other countries[37]. Such emerging clinical evidence emphasizes the need for further investigation into olfactory ACE2 expression via scRNA-seq and other modalities.

Taken together, these scRNA-seq analyses and triangulation to literature synthesis confirm that type II pneumocytes are a likely target of SARS-CoV-2 infection while also highlighting club cells, ciliated cells, and olfactory epithelial cells as additional potential sites of infection.

## Tongue keratinocytes and mature small intestinal enterocytes are potential targets of SARS-CoV-2

We then classified 136 gastrointestinal cell types from nine scRNA-seq studies based on their expression of and literature associations to ACE2 (**Figure 4A**). These studies encompassed samples from the upper, mid, and lower GI tracts including tongue, esophagus, stomach, small intestine, and colon[19,38–42].

This analysis highlights a robust expression of ACE2 in tongue keratinocytes that has not been strongly documented in the literature, as evidenced by the weak local context score between ACE2 and keratinocytes (**Figure 4B**). In fact, we found no previous reports of ACE2 expression in keratinocytes and only one recent report suggesting ACE2 expression in the human tongue based on a combination of bulk RNA-seq and a scRNA-seq dataset which has not been made publicly accessible[43]. We propose that a subset of ACE2$^+$ tongue keratinocytes may serve as a novel site of SARS-CoV-2 entry and highlight the need to generate additional gene and protein expression data from human tongue samples to further evaluate this hypothesis. Emerging reports of loss of taste (dysgeusia) in otherwise non-symptomatic COVID-19 patients may warrant further study of the tongue in this pathology[44].

We also found that ACE2 is highly expressed in both human and murine small intestinal enterocytes, confirming an association which has been moderately appreciated in literature, as indicated by our literature derived local score between ACE2 and enterocytes. However, to our knowledge, the transcriptional heterogeneity of ACE2 among enterocyte populations has never been explored. In this context, we found that ACE2 shows an increase in expression correlated with the maturation of murine small intestinal enterocytes, with minimal expression in stem cells and transit amplifying cells in contrast to most robust expression in mature enterocytes (**Figure 4**). To the best of our knowledge, this is the first demonstration that ACE2 expression synchronously increases over the course of enterocyte maturation. The recognition of such intra-tissue heterogeneity is necessary to specify the cell types which are most likely responsible for the proposed fecal-oral transmission of COVID-19[12].

**COVID-19, SARS, MERS and HCoV-229E receptors share a transcriptional signature correlated to maturation of small intestinal enterocytes**

To determine whether the maturation-correlated expression pattern is unique to ACE2, we computed cosine similarities between the ACE2 gene expression vector (CP10K values in ~6,000 small intestinal enterocytes) and that of the ~15,700 other genes detected in this study (**Figure 5A**). For this analysis, the vector space is constituted of the individual cells as the dimensions using the gene expression values to construct the vectors (see **Methods**). Interestingly, we found that ANPEP, the established entry receptor for HCoV-229E, showed the third highest cosine similarity to ACE2 (**Figure 5B**). Further, DPP4 – the entry receptor for MERS coronavirus – is also among the top 1% of similarly expressed genes by this metric (**Figure 5B**). We confirmed that both of these genes do indeed show a maturation-correlated transcriptional pattern similar to that of ACE2 (**Figure 5C-D**), highlighting an unexpected shared pattern of transcriptional heterogeneity among known coronavirus receptors in a cell population which may be relevant for viral transmission.

We then asked whether this shared pattern of transcriptional heterogeneity among coronavirus receptors is observed in the human small intestine. Indeed, among all enterocytes from a human scRNA-seq study, both ANPEP and DPP4 were among the top 1% of genes with similar expression vectors to that of ACE2 (**Figure S22A-B**). We independently validated this observation by computing gene expression correlations from bulk RNA-sequencing of human small intestine samples from GTEx (n = 187), which similarly revealed that DPP4 and ANPEP are among the top 1% of correlated genes to ACE2 (**Figure S22C**). In fact, among all ~18,500 genes mean expression > 1 TPM in GTEx small intestine samples, DPP4 shows the second highest correlation to ACE2 (r = 0.95).

To our knowledge, this is the first demonstration that all known coronavirus entry receptors display highly coordinated and maturation-correlated transcriptional expression patterns in intestinal epithelial cells. We propose that the requisite interaction with human proteins displaying a tightly defined expression gradient on apical surfaces of epithelial cells, which is shared among known coronavirus strains, may have fundamental implications for understanding the evolution, lifecycle, and/or transmission patterns of this family of viruses.

## Discussion

Recent advances in scRNA-seq are empowering us to study tissue and cellular transcriptomes at previously unprecedented resolutions. Several single-cell RNA sequencing based efforts such as the Human Cell Atlas are underway to catalog gene expression across tissues and cell types, and the raw data from many of these studies are available on public platforms such as the Broad Institute Single Cell Portal[45] and Gene Expression Omnibus (GEO). Analyses of these datasets are of interest to a wide range of researchers but currently prove challenging for all but a few due to the need for specialized workflows and computing infrastructures. Consequently, the widespread use of this data for biomedical research is hampered, an issue which is particularly evident in the face of public health crises like the ongoing COVID-19 pandemic. To address this unmet need, the nferX platform Single Cell resource enables the rapid and interactive analysis of the continually growing scRNAseq datasets by specialists and non-specialists alike. Furthermore, the seamless triangulation of scRNA-seq insights with global and local scores derived from the synthesis of accessible biomedical literature creates a truly first-in-class resource.

By making the resource available to all academic researchers, we enable scientists to not only dive deeper into insights that are aligned with existing knowledge but also to prioritize the novel insights which warrant further experimental validation. Looking forward, we plan to automate the integration of the rapidly growing number of scRNA-seq studies so that access to the entire world's knowledge of single cell transcriptomes is just one click away for any researcher. As we do so, we encourage interactive feedback from the scientific community so that this platform can evolve to optimally support the research needs across the biomedical ecosystem, beyond the COVID-19 focus on the current study.

Combined with our analyses of bulk RNA-seq, IHC, and proteomics datasets, our characterization of the known human coronavirus receptors (ACE2, DPP4, ANPEP) using the nferX platform Single Cell resource represents the most comprehensive molecular fingerprint of host factors determining coronavirus infections including COVID-19. While this serves as a primer of the deep profiling that is made possible with this resource, we also identified several interesting aspects of coronavirus receptor biology which warrant further experimental follow-up.

We identified tongue keratinocytes and olfactory epithelia as novel ACE2-expressing cell populations and thus as important potential sites of SARS-CoV-2 infection. This molecular fingerprint is a striking correlate to emerging clinical reports of dysgeusia[44] and anosmia[37] in COVID-19 patients, which strongly implicate the gustatory and olfactory systems in SARS-CoV-2 pathogenesis and human-to-human transmission. Tongue epithelial cells have also previously been shown to uptake Epstein-Barr virus[46], and importantly a recent study found that ACE2 is appreciably expressed in tongue based on a small number of non-tumor bulk RNA-seq samples from TCGA[43]. This study further showed by scRNA-seq that ACE2 expression is observed in a subset of the human tongue (but not other oral mucosal) epithelial cells, albeit in only ~0.5% of the recovered epithelial population. This data has unfortunately not been released for public consumption but certainly does provide preliminary support for our finding, particularly as the listed set of cluster-defining genes for this population (SFN, KRT6A, KRT10) is consistent with the tongue keratinocyte identify from the *Tabula Muris* data set (**Figure S23**). We thus emphasize the imminent need for further generation of multi-omic expression data from large numbers of healthy and diseased human tongue samples drawn from a cohort of wide demographic representation.

We also observed that expression of ACE2 and other coronavirus receptors is intimately linked to the maturation status of small intestinal enterocytes, pinpointing the more mature subsets as the most likely cells to harbor SARS-CoV-2 virus. This finding amplifies the potential for fecal-oral transmission of COVID-19[10–12] and should motivate further experimental validation to determine whether monitoring of fecal viral loads should be considered clinically for diagnostic or prognostic purposes.

We further found that this transcriptional mirroring of coronavirus entry receptors was not unique to small intestine but rather also strongly present among renal proximal tubule epithelial cells, where ACE2, DPP4, and ANPEP expression tends to be observed in the same cellular subsets. These observations suggest the existence of a transcriptional network spanning tissues and cell types which may drive and/or regulate coronavirus receptor expression. The question of whether coronaviruses have evolved to exploit such a network may be relevant to pursue, particularly given that downregulation of ACE2 by SARS-CoV has been reported previously and is associated with poor clinical outcomes[31,47]. Perhaps other coronaviruses can similarly modulate the expression of their entry receptors to impact the clinical course of the induced disease.

The emerging picture of the coronavirus life cycle appears to be intricately interwoven with many proteins beyond the primary host receptors. For instance, a recent structural complex of the SARS-CoV-2 spike protein with ACE2 identified SLC6A19 as an interaction partner of ACE2[48]. Further, spike proteins from some coronaviruses can interact with CEACAM1[49] and sialylated glycans similar to influenza hemagglutinin[50] as host receptors. Future studies are likely to highlight several other proteins and glycans that constitute the "interactome" of the coronavirus proteome. Understanding the expression profiles of the interactome across tissues will provide systems level insights on the cellular dynamics of the functional partners and the regulatory machinery of the host receptor proteins. Like in the current study, the nferX platform will be an excellent resource for unraveling the purported interaction partners for coronavirus receptors and profiling their expression across different tissues and cells constituting the human body.

Overall, this study evidences the utility of an integrative data science platform to enable rapid and high-throughput analysis of publicly available data to generate relevant biological insights and scientific hypotheses. We hope that by making our biomedical knowledge synthesis-augmented single cell platform publicly accessible, we help empower the research community to advance our understanding of the world's most pressing biomedical challenges such as COVID-19.

## Methods

**Unstructured biomedical knowledge synthesis and triangulation capabilities**

In order to capture biomedical literature based associations, the nferX platform defines two scores: a "local score" and a "global score", as described previously[51]. Briefly, the local score represents a traditional natural language processing technique which captures the strength of association between two concepts in a selected corpus of biomedical literature based on the frequency of their co-occurrence normalized by the frequency of each individual concept throughout the corpus. A higher local score between Concept X and Concept Y indicates that these concepts are frequently mentioned in close proximity to each other more frequently than would be expected by chance. The global score, on the other hand, is based on the neural network renaissance that has recently taken place in the Natural Language Processing (NLP) field. To compute global scores, all tokens (e.g. words and phrases) are projected in a high-dimensional vector space of word embeddings. These vectors serve to represent the "neighborhood" of concepts which occur around a given concept. The cosine distance between any two vectors measures the similarity of these neighborhoods and is the basis for our global score metric, where concepts which are more similar in this vector space have a higher global score.

While the global scores in this work are computed in the embedding space of word2vec model, it can also be computed in the embedding space of any deep learning model including recent transformer-based models like BERT[52]. These may have complementary benefits to word2vec embeddings since the embeddings are context sensitive having different vectors for different sentence contexts. However, despite the context sensitive nature of BERT embeddings a global score computation for a phrase may still be of value given the score is computed across sentence embeddings capturing the context sensitive nature of those phrases.

From a visualization perspective, the local score and global score ("Signals") are represented in the platform using bubbles where bubble size corresponds to the local score and color intensity corresponds to the global score. This allows users to rapidly determine the strength of association between any two concepts throughout biomedical literature. We consider concepts which show both high local and global scores to be "concordant" and have found that these typically recapitulate well-known associations.

The nferX platform also supports a logical thought engine that enables AND (conjunction), OR (disjunction), and NOT (negation) queries - the universal logic gates. This engine is referred to as "dynamic adjacency" and leverages a highly distributed main memory approach that allows the computation of local scores for any type of logical query in real time. Fundamentally, this system allows a user to extract all 100-word fragments of text which meet the specified logical query. We then calculate local scores for all other tokens occurring within these fragments, which quantifies the likelihood (i.e. odds) of each token occurring this frequently within these textual fragments by chance.

The platform further leverages statistical inference to calculate "enrichments" based on structured data, thus enabling real-time triangulation of signals from the unstructured biomedical knowledge graph various other structured databases (e.g. curated ontologies, RNA-sequencing datasets, human genetic associations, protein-protein interactions). This facilitates unbiased hypothesis-free learning and faster pattern recognition, and it allows users to more holistically determine the veracity of concept associations. Finally, the platform allows the user to identify and further examine the documents and textual fragments from which the knowledge synthesis signals are derived using the Documents and Signals applications.

**Association Scores**

If we have an automated method that, given a corpus (consisting of text and other structured and semi-structured data as is often the case with biomedical data), comes up with a *strength of association score* between a query entity and all the tokens or entities present in the corpus, then that score can be used to obtain a ranking of tokens or entities related to the query. There have been other motivations for ranking the association strength of tokens/entities (e.g. for use in picking among possible tokens in speech recognition or Optical Character Recognition - OCR).

Generally, association scores are based in some way on the co-occurrences of the tokens (or referents to the entities) in the text within small windows of text. Co-occurrences have been studied in linguistics/NLP since at least Firth's maxim that "a word is known by the company it keeps." One popular traditional measure for association strength between tokens in text is pointwise mutual information, or PMI[53], which we consider in several association scores.

**Measures of association**

Formally, for a given corpus, an *association score* is some real-valued function *S(q, t)* where *q* is a query token/entity and *t* is another token/entity. We discussed association scores as if they were symmetric above, but for some convenience later on our formal notion is asymmetric: there's a query *q* and a token *t*. In particular, we later extend *q* to logical combinations of tokens/entities. The association score need not be symmetric (for logical queries it cannot be as *t* is still restricted to single tokens/entities) though it often is when *q* and *t* are both single tokens/entities.

*Context of q* - All the measures involve the notion of the "context" of the query *q*. The context of *q* are those locations in the corpus deemed to be "near" to *q*. For single token queries, follow the typical approach of defining context as those locations in the corpus that are within some fixed number of words *w* (the window size *w* is a tunable parameter) from an occurrence of *q* in the corpus. The dynamic adjacency engine generalizes this notion of context in a natural way to logical queries; the context for a logical *q* is a certain set of fixed-length fragments.

*Co-occurrences* - This is just the number of times *t* appears in the context of *q*.

*Traditional PMI* - This is $\log(p(t \mid q) / p(t))$. Here p(*t* | *q*) is the number of times t occurs in the context of q (ie co-occurrences of t and q) divided by the total length of all q contexts in the corpus, whereas p(*t*) is the number of occurrences of *t* in the entire corpus, divided by the corpus length.

*Word2vec cosine distance* - The popular word2vec algorithm [5] generates a vector (we use 300-dimensional vector representation) for each token in a corpus. The purpose of these vectors is usually to be used as features in downstream NLP tasks. But they can also be used for similarity. The original paper validates the vectors by testing them on word similarity tasks: the association score is the cosine between the vector for *q* and the vector for *t*. This score only applies to single-token *q*.

*Exponential mask PMI (ExpPMI)* - This is our first new proposed score. PMI treats every position in a binary way: it's either in the context of *q* or not. With a window size of say 50, a token which appears 3 words from a query *q* and a token which appears 45 words from a query *q* are treated the same. We thought it might be useful to consider a measure which distinguishes positions in the context based on the number of words away that position is from an occurrence of *q*. We did this by weighting the positions in the context by some weight between 0 and 1. Our weighting is based on an exponential decay (which has some nice properties especially when we extend to the case of logical queries).

*Local score* - This is another new proposed score. We find that PMI and ExpPMI can vary a lot for small samples (i.e. small numbers of co-occurrences, occurrences). The Local Score is *log(coocc) * sigmoid(PMI - 0.5)*, constructed to correct for this; we found that this formula too works well empirically.

*Exponential mask local score (ExpLocalScore)* - We apply both modifications together: the exponential mask score is *log(weighted_coocc) * sigmoid(expPMI - 0.5)*. Here *weighted_coocc* is the sum of the weights of the positions of the corpus

Evaluation of association scores are further described in the Supplementary Information.

**Single-cell RNA-seq analysis platform**

The objective of the single cell platform is to enable dynamic visualization and analysis of single cell RNA-sequencing data. Currently, there are over 30 scRNA-seq studies available for analysis in the Single Cell app including studies from human donors/patients covering tissues such as adipose tissue, blood, bone marrow, colon, esophagus, liver, lung, kidney, ovary, nasal epithelium, pancreas, placenta, prostate, retina, small intestine, and spleen. Because no pan-tissue reference dataset yet exists for humans, we have manually selected individual studies to maximally cover the set of human tissues. In some cases, these studies contain cells from both healthy donors and patients affected by a specified pathology such as ulcerative colitis (colon) or asthma (lung). There are also a number of murine scRNA-seq studies covering tissues including adipose tissue, airway epithelium, blood, bone marrow, brain, breast, colon, heart, kidney, liver, lung, ovary, pancreas, placenta, prostate, skeletal muscle, skin, spleen, stomach, small intestine, testis, thymus, tongue, trachea, urinary bladder, uterus, and vasculature. Note that two of these murine studies (Tabula Muris and Mouse Cell Atlas) include ~20 tissues each.

**Single-cell data processing pipeline**
For each study, a counts matrix was downloaded from a public data repository such as the Gene Expression Omnibus (GEO) or the Broad Institute Single Cell Portal (**Table S1**). Note that this data has not been re-processed from the raw sequencing output, and so it is likely that alignment and quantification of gene expression was performed using different tools for different studies. In some cases, multiple complementary datasets have been generated from a single publication. In these cases, we have generated separate entries in the Single Cell platform.

While counts matrices have been generated using different technologies (e.g. Drop-Seq, 10x Genomics, etc.) and different alignment/pre-processing pipelines, all counts matrices were scaled such that each cell contains a total of 10,000 scaled counts (i.e. the sum of expression values for all genes equals 10,000 in each individual cell). All data were uniformly processed using the Seurat v3 package[54]. In short, this pipeline involves the following steps. First, we identify 2000 variable genes across the given dataset and then perform linear dimensionality reduction by principal component analysis (PCA). Using the set of principal components which contribute >80% of variance across the dataset, we then do the following: (i) perform graph-based clustering to identify groups of cells with similar expression profiles (Louvain clustering), (ii) compute UMAP and tSNE coordinates for each individual cell (used for data visualization) and (iii) annotate cell clusters. Note that the three human pancreatic datasets (GSE81076, GSE85241, GSE86469) were integrated together in a shared multi-dimensional space using CCA (Canonical Correlation Analysis) and the integration method in the Seurat v3 package[54]. Cell clustering and computation of dimensionality reduction coordinates were performed on this integrated dataset.

**Cell cluster annotation**

In cases where publicly deposited counts matrices are accompanied by author-assigned annotations for individual cells or clusters, we have retained these cell annotations for display in the platform and accompanying analyses. For any study which was not accompanied by a metadata file containing cluster annotations, we have manually labeled clusters based on sets of canonical "cluster-defining genes." In these cases, we have attempted to leverage annotations and descriptions of gene expression patterns described by study authors in the manuscript text and figures corresponding to the data being analyzed.

**Metrics to Summarize Cluster-Level Gene Expression**

The platform allows users to query any gene in any selected study. The corresponding data is displayed in commonly employed formats including a series of violin plots and as a set of dimensionality reduction plots. Expression is summarized by listing the percent of cells expressing Gene *G* in each annotated cluster and the mean expression of Gene *G* in each cluster. To measure the specificity of Gene *G* expression to each Cluster *C*, we compute a cohen's D value which assesses the effect size between the mean expression of Gene *G* in cluster *C* and the mean expression of Gene *G* in all other clusters. Specifically, the cohen's D formula is given as follows: $(Mean_C - Mean_A)/(sqrt(StDev_C^2 + StDev_A^2))$, where *C* represents the cluster of interest and *A* represents the complement of *C* (i.e. all other cell clusters). Note that this is functionally similar to the computation of paired fold change values and p-values between clusters which is frequently used to identify cluster-defining genes.

**Gene-Gene Cosine Similarity**

Within the platform, we support the run-time computation of cosine similarity (i.e. 1 - cosine distance) between the queried gene and all other genes. This provides a measure of expression similarity across cells and can be used to identify co-regulated and co-expressed genes. Specifically, to perform this computation, we construct a "gene expression vector" for each gene *G*. This corresponds to the set of CP10K values for gene *G* in each individual cell from the selected populations in the selected study.

**Profiling Expression of Coronavirus Receptors in Single-cell Datasets**

For each single-cell dataset, we examined the expression of *ACE2*, *TMPRSS2*, *ANPEP*, and *DPP4*. We generally considered a cell population to potentially express a gene if at least 5% of cells from that cluster showed non-zero expression of this gene. For each dataset, we show a figure which includes a UMAP dimensionality reduction plot colored by annotated cell type along with identical plots colored by the expression level of each coronavirus receptor in all individual cells. In some cases, we also show violin plots from the platform which automatically integrate literature-derived insights to highlight whether there exist textual associations between the queried gene and the tissue/cell types identified in the selected study.

## Acknowledgments

The authors thank Murali Aravamudan, Peter Lebowitz, Ajit Rajasekharan, and Mathai Mammen for their insightful reviews and feedback on our research. We also express our gratitude to Patrick Lenehan, Saurav Kumar Verma, Travis Hughes, and Vishy Thiagarajan who helped develop some of the scientific tools that were leveraged for this study.

## Figure Legends

**Figure 1. Knowledge synthesis and the nferX Single Cell resource**. **(A)** Knowledge synthesis: capturing association between concepts from over 100 million documents. Schematic shows the work-flow for generating literature-derived associations between phrases. Local score and global score are defined and the types of literature-derived associations are shown for combinations of high and low local and global scores. **(B)** Datasets enabling knowledge synthesis-powered scRNA-seq analysis platform. Single Cell RNAseq data was obtained from over 30 publicly available human and mouse single cell RNA-seq datasets. Bulk RNA-seq data was obtained from Gene Expression Omnibus (GEO) and the Genotype Tissue Expression (GTEx) project portal. Protein-level expression of coronavirus receptors was assessed using a collection of immunohistochemistry (IHC) images and tissue proteomics datasets from the Human Protein Atlas and the Human Proteome Map. Literature-derived association scores are obtained from over 100 million biomedical documents **(C)** Tissues and cell-types identified by one or more modalities to express ACE2, the putative receptor of SARS-CoV-2 spike protein.

**Figure 2. Triangulation of knowledge synthesis with ACE2 expression profile by scRNA-seq across cells and tissues.** Scatterplot shows comparison of percentage of cells with non-zero expression (x-axis) against literature-derived associations: local score (y-axis and size of circles) and global score (transparency of circles). Data includes more than 800 cell populations from over 25 murine and human tissues comprising over 1 million cells.

**Figure 3. Triangulation of ACE2 expression in the respiratory tract with literature-derived insights. (A)** Schematic representation of the respiratory system highlighting key cell types from the nasal cavity, airway and alveoli. Scatterplot shows comparison of percentage of cells with non-zero expression (x-axis) from eight single cell studies against literature-derived associations: local score (y-axis and size of circles) and global score (transparency of circles). **(B)** Assessing literature-based and scRNAseq-based associations between ACE2 and respiratory tract cells. On the left, the dimensionality reduction plots show different cell populations associated with lung and olfactory epithelium. On the right, violin plots show the distribution of ACE2 expression levels in selected populations with non-zero expression. The cell-types and the literature-derived local and global associations scores are shown.

**Figure 4. Triangulation of ACE2 expression in the gastrointestinal (GI) tract with literature-derived insights. (A)** Schematic representation of the GI tract highlighting key cell types. Scatterplot shows comparison of percentage of cells with non-zero expression (x-axis) from nine single cell studies against literature-derived associations: local score (y-axis and size of circles) and global score (transparency of circles). **(B)** Assessing literature-based and scRNAseq-based association between ACE2 and tongue keratinocytes. Violin plot shows distributions of ACE2 expression in keratinocytes. The literature-derived local and global associations between ACE2 and keratinocytes are shown. (**C**) ACE2 transcriptional expression is correlated to enterocyte maturation. Violin plots show distribution of ACE2 expression levels in enterocytes at different stages of differentiation.

**Figure 5. Coronavirus receptors share a transcriptional signature correlated to maturation of small intestinal enterocytes.** (**A**) Distribution of cosine distances between the 'gene expression vectors' of ACE2 and all genes in a scRNA-seq study of the murine small intestine. The gene expression vector corresponds to the set of CP10K values for a given gene in each individual cell from the selected populations in the selected study. (**B**) Genes similar to ACE2 (cosine similarity > 0.4) sorted by literature-derivation association. (**C**) Transcriptional expression of ANPEP correlated to enterocyte maturation in murine small intestine. Violin plots show distribution of ANPEP expression levels in enterocytes at different stages of differentiation. (**D**) Transcriptional expression of DPP4 correlated to enterocyte maturation in murine small intestine. Violin plots show distribution of DPP4 expression levels in enterocytes at different stages of differentiation.

## A. Knowledge synthesis: capturing association between concepts from over 100 million documents

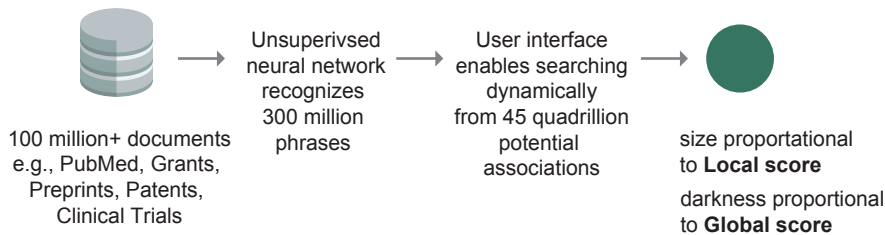
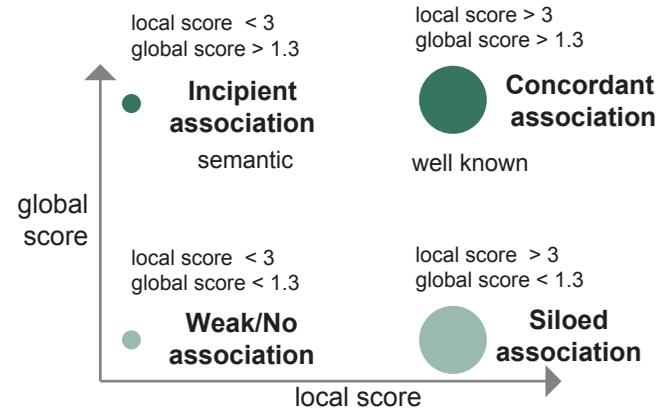

**Local score** measures how frequently two tokens are found within each other's local context in a particular corpus, normalized by the occurrences of those tokens in that corpus. Derivative of point-wise mutual information.

**Global score** measures the similarity between the word vectors corresponding to two tokens, normalized by a control token collection. Word2Vec at scale.

## B. Knowledge synthesis powered scRNAseq platform

~1 million cells
25 tissues
30 studies

Identify cell populations
Identify cluster-defining genes
Leverage machine learning based literature associations

>100 million biomedical documents (Includes COVID-19 open research dataset)

e.g., PubMed, Grants, Preprints, Patents, Clinical Trials

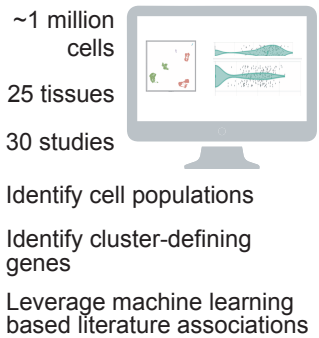
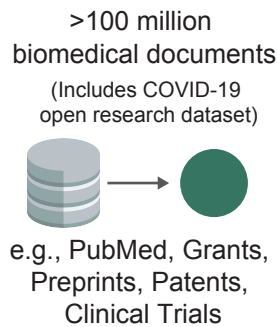

Bulk RNAseq — 262,152 RNA samples; GEO (6,552 studies), GTEx (54 studies)
Proteomics — Mass spec >25 healthy tissues; Human Protein Atlas, Human Proteome Map
IHC — Antibody-based quantification in >40 healthy tissues; Human Protein Atlas

## C. Expression profiling of ACE2 Putative receptor of SARS-CoV-2

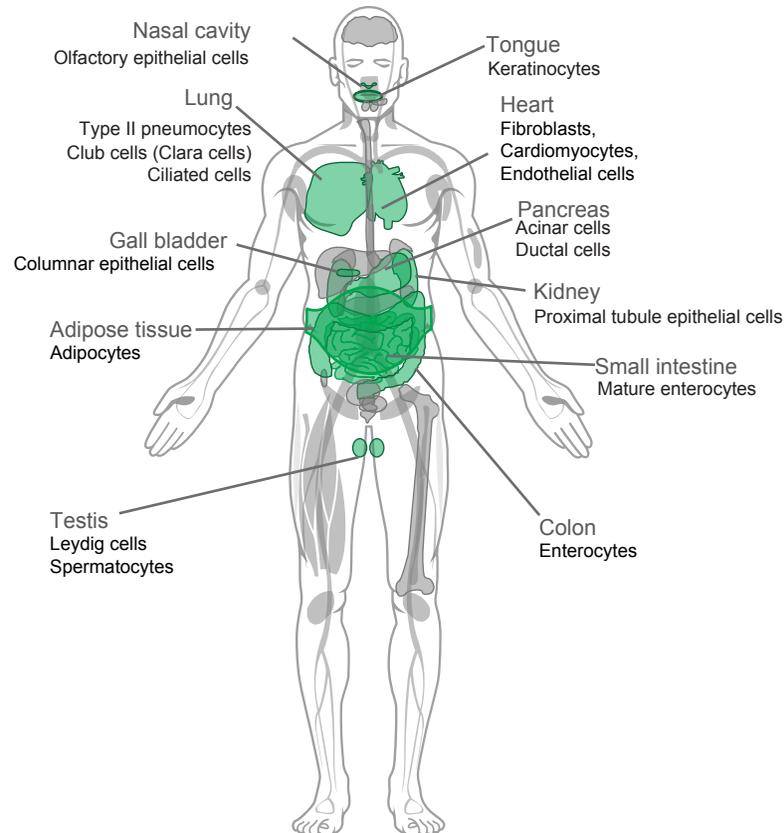

- Nasal cavity — Olfactory epithelial cells
- Tongue — Keratinocytes
- Lung — Type II pneumocytes, Club cells (Clara cells), Ciliated cells
- Heart — Fibroblasts, Cardiomyocytes, Endothelial cells
- Gall bladder — Columnar epithelial cells
- Pancreas — Acinar cells, Ductal cells
- Adipose tissue — Adipocytes
- Kidney — Proximal tubule epithelial cells
- Small intestine — Mature enterocytes
- Testis — Leydig cells, Spermatocytes
- Colon — Enterocytes

**Figure 1**

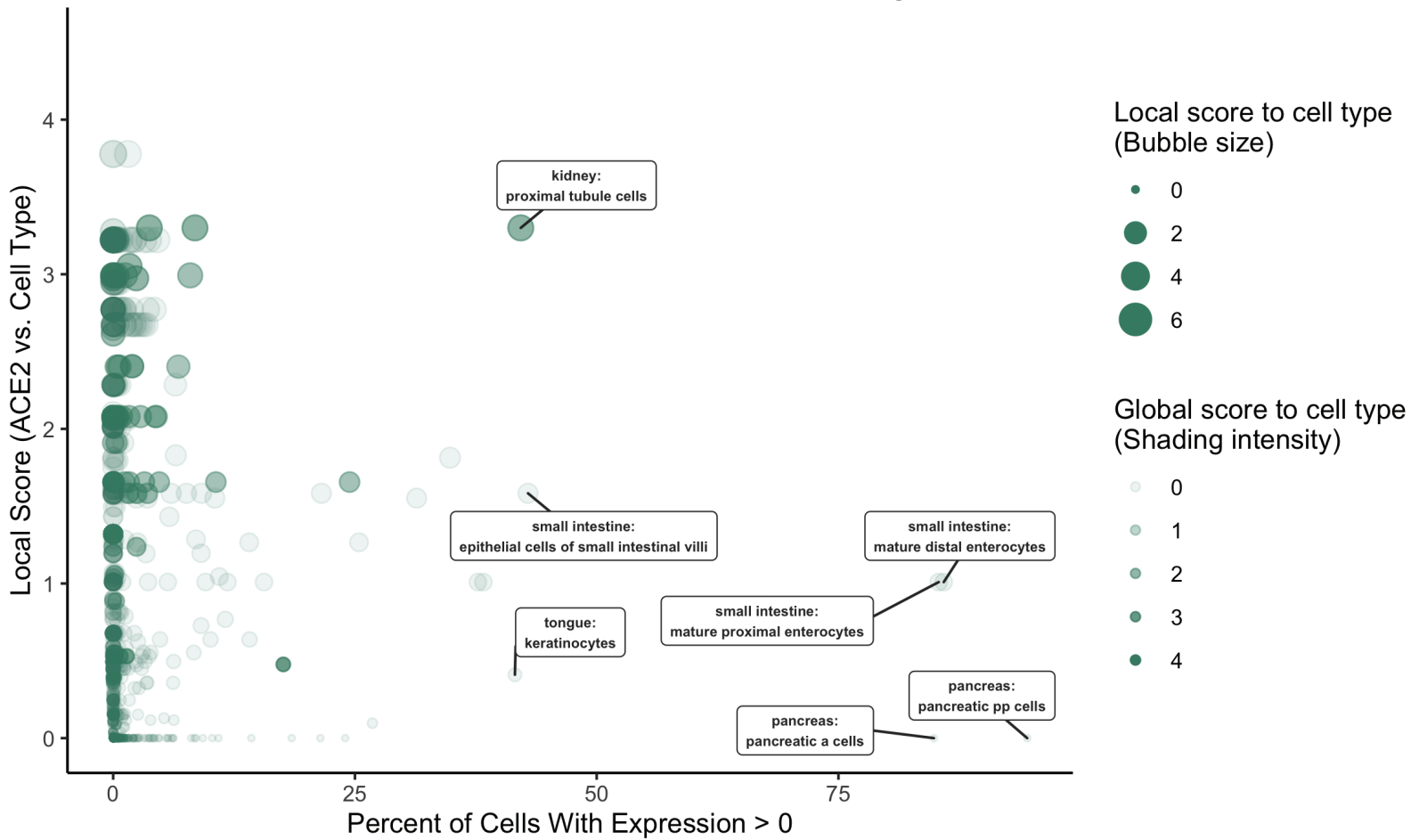

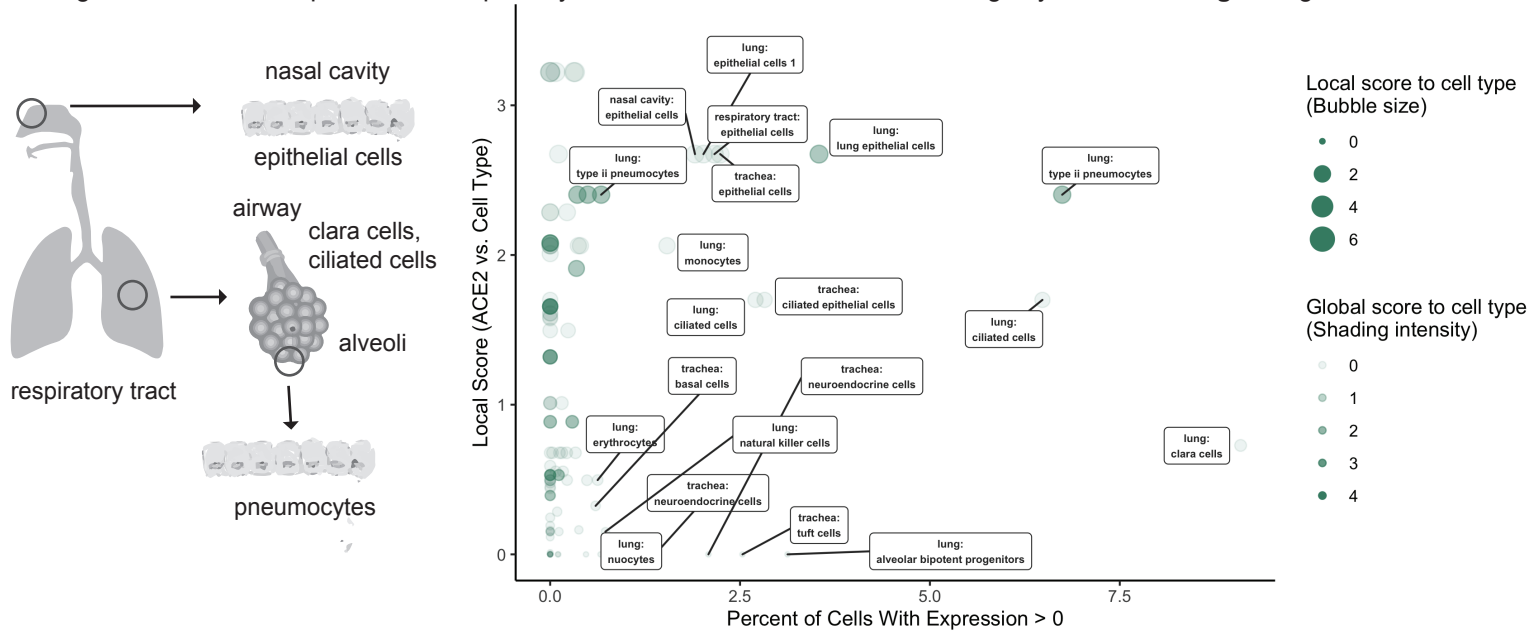

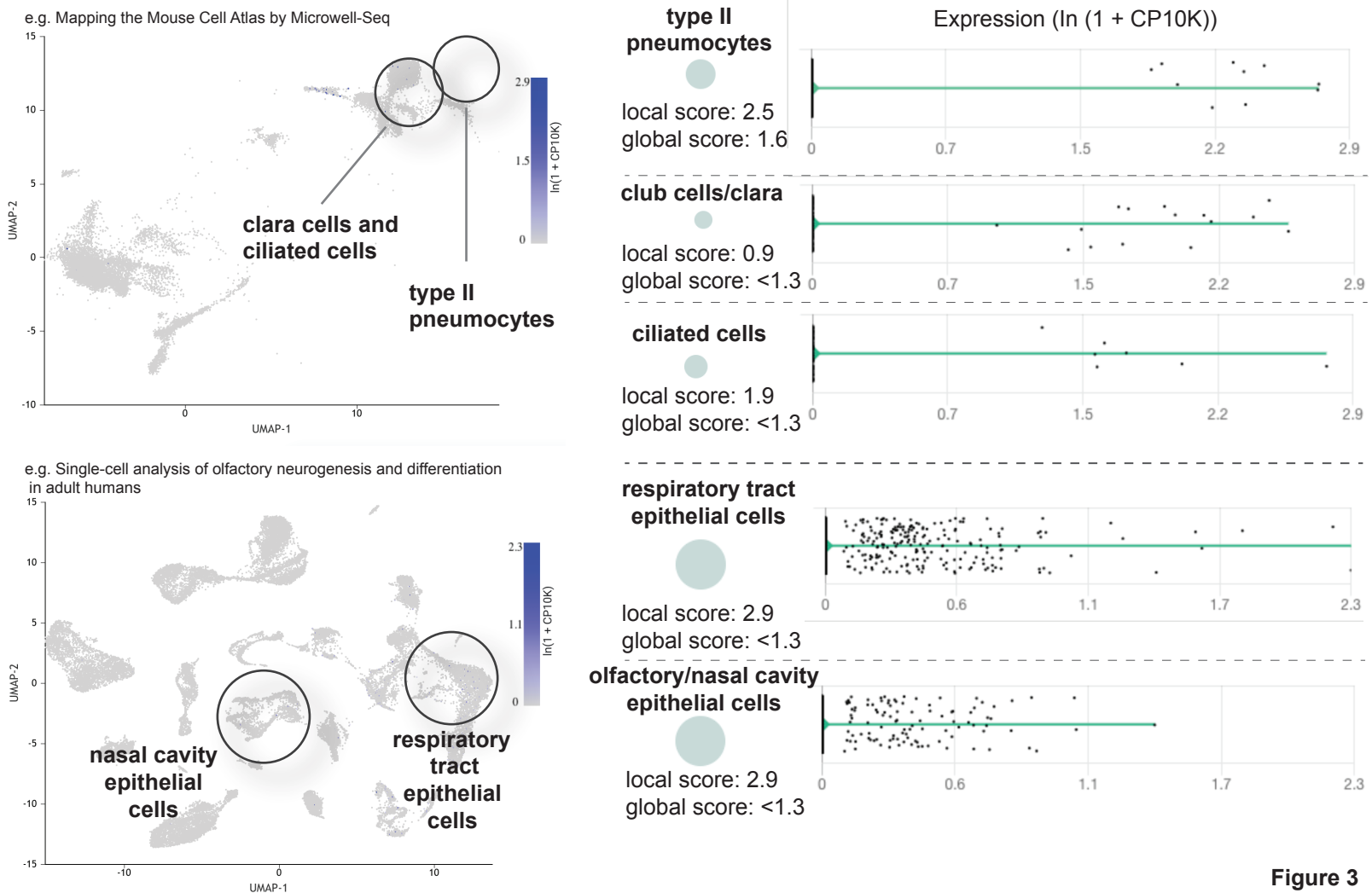

Figure 3

## A ACE2 expressing cell populations in the gastrointestinal tract

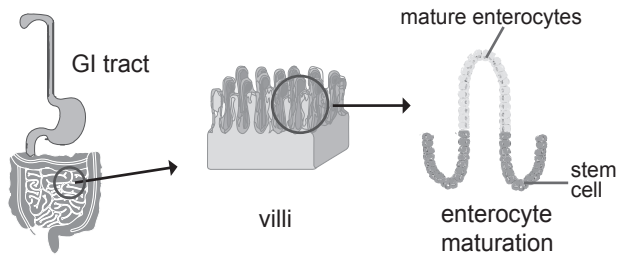

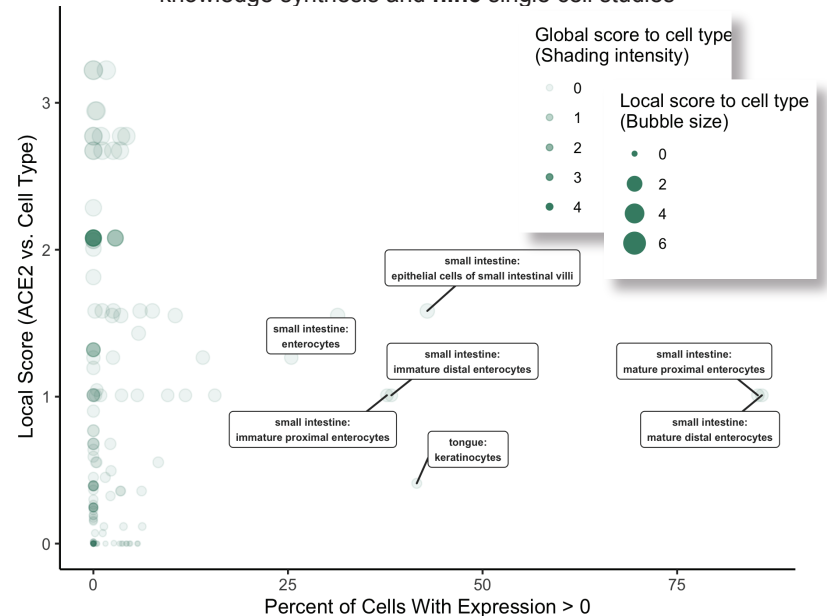

## B ACE2 expression in tongue keratinocytes

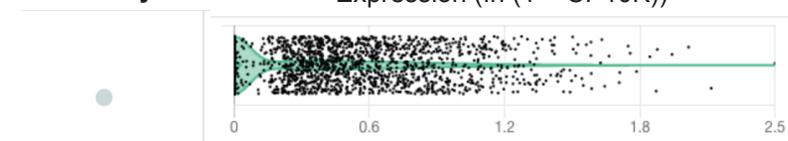

e.g. Single-cell transcriptomics of 20 mouse organs creates a Tabula Muris.

## C transcriptional signature correlated to enterocyte maturation

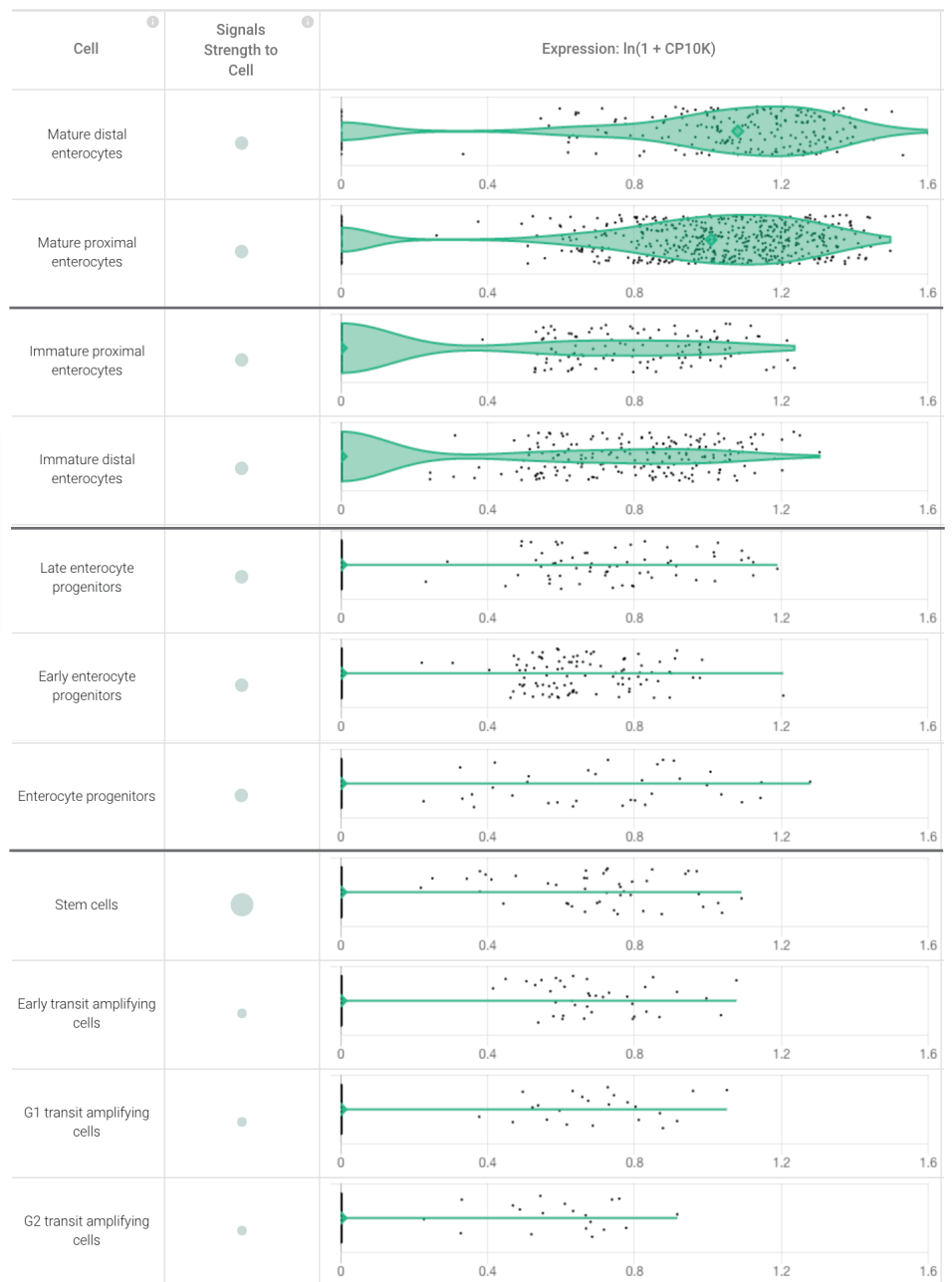

**Figure 4**

# Expression of all known coronavirus receptors synchronously increase with enterocyte maturation

**A** 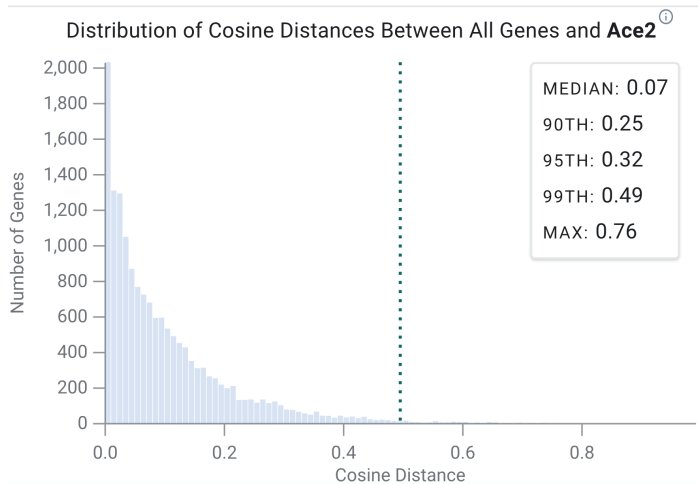

**B** 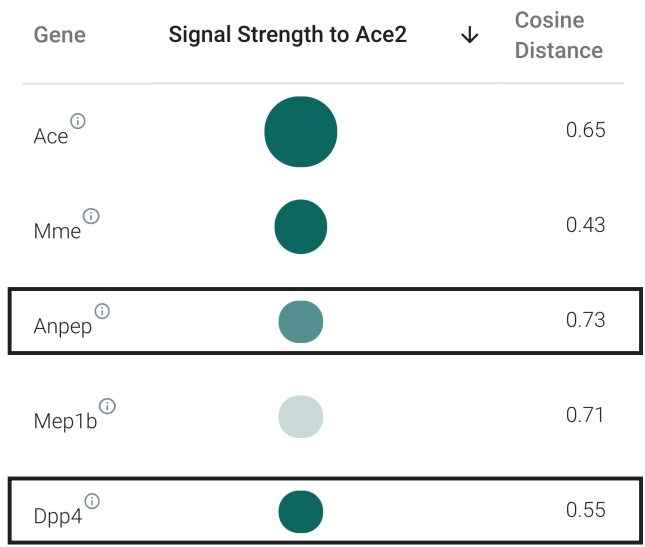

**C** 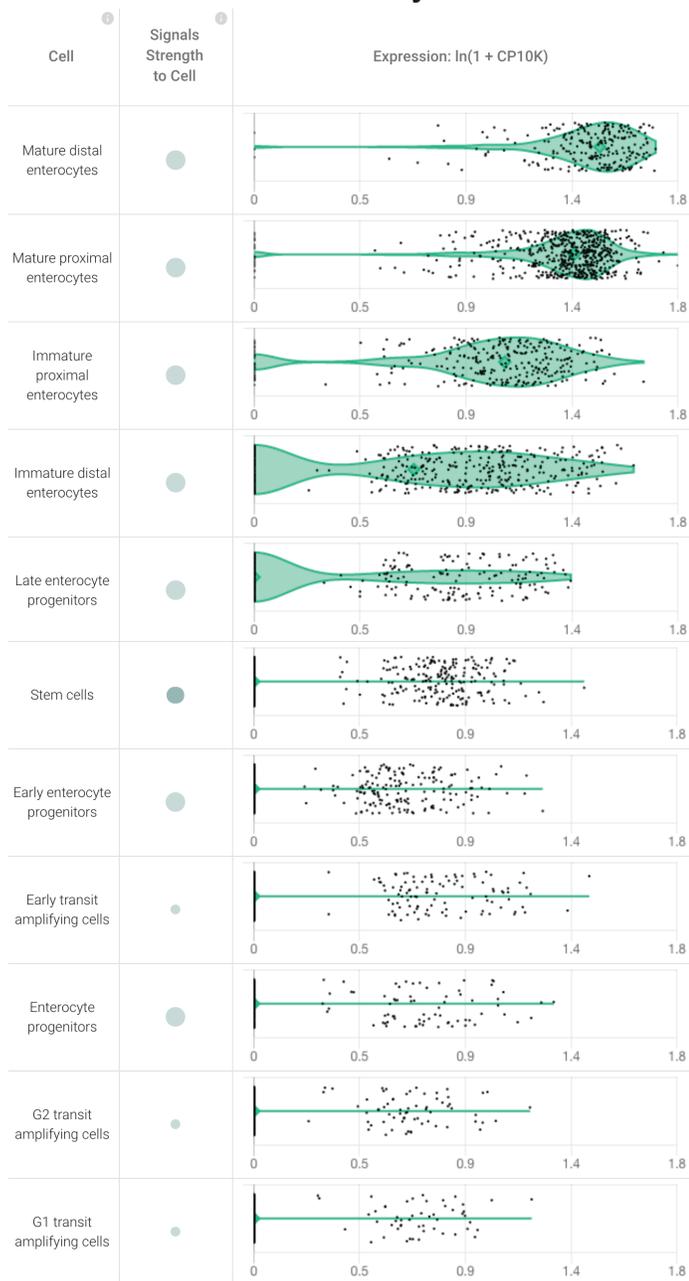

**D** 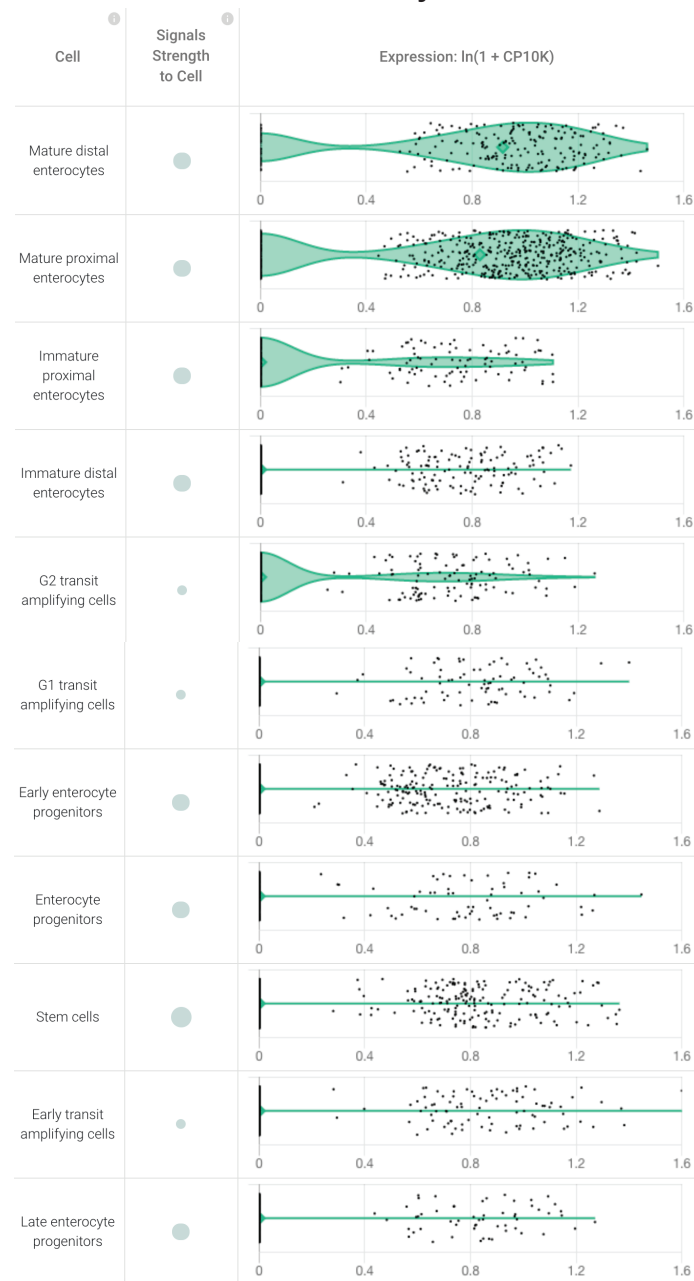

Figure 5